\documentclass[12pt]{iopart}

\usepackage{iopams} 
\usepackage{epsf}
\usepackage{subfigure}
\usepackage{graphicx} 
\begin{document}

\title{Universality class of isotropic on-lattice Eden clusters}

\author{L. R. Paiva and S. C. Ferreira Jr.\footnote[3]{To whom correspondence should be addressed (silviojr@ufv.br)}}

\address{Departamento de F\'{\i}sica, Universidade Federal Vi\c{c}osa, 36571-000, Vi\c{c}osa, MG, Brazil}

\begin{abstract}
The shape of large on-lattice Eden clusters grown from a single seed is ruled by the underlying lattice anisotropy. This is reflected on the linear growth with time of the interface width ($w\sim t$), in contrast with the KPZ universality class ($w\sim t^{1/3}$) observed when the Eden model is grown on flat substrates. We propose an extended Eden model, in which the growth probability has a power law dependence with  the number of occupied nearest neighbors. Large scale simulations ($N\gtrsim 4\times 10^9$ particles) were used to determine the time evolution of $w$. We found that a suitable choice of the power exponent removes the lattice-induced cluster anisotropy and provides a growth exponent in very good agreement with the KPZ universality class. Also, the present model corroborates the results found in off-lattice simulations, in which the center of mass fluctuations are considered in the interface scaling analysis.
\end{abstract}


\submitto{\it J. Phys. A: Math. Gen.}

\maketitle

\section{Introduction}

Since the interfaces are ubiquitous in nature, very intensive efforts are devoted to the description of their physical properties \cite{Barabasibook}. Models where discrete unities are arranged in regular lattices constitute very useful tools due to their easy computer implementation. However, on-lattice growth models are frequently very sensitive to the intrinsic lattice anisotropy. Such a dependence is evident for clusters grown from a single particle (a seed) as, for example, the classical diffusion-limited aggregation (DLA) \cite{Tolman,Goold} and Eden \cite{Zabolitzky,Batchelor} models. Consequently, the interface scaling analysis in these models is much less frequent than that for the growth from flat surfaces. However, the interface scaling in several experiments with radial symmetry has been considered along the past decade. Examples include the dynamics of several types of tumors \cite{Bru1998,Bru2003,Bru2004}, the plant callus evolution \cite{Galeano}, and grain-grain displacement in Hele-Shaw cells \cite{Couto}, among others. In this scenario, the elaboration of on-lattice models without the undesirable effects of anisotropy becomes worthy. 

The Eden model \cite{Eden} is the simplest discrete model, in which the anisotropy is determinant. Its original version was studied on a square lattice where the sites can assume two values $1$ or $0$, representing occupied or empty sites, respectively. The   simulations start with a single particle at the center of the lattice and the growth rules are the following: at each step a peripheral site (an occupied site with at least one empty nearest-neighbor) is chosen at random and one of its empty nearest-neighbors is selected with equal probability and then occupied. Variants of the Eden rules were studied, and the original version is commonly called Eden B \cite{Meakinbook}. 

Despite its simplicity, the Eden model produces clusters with nontrivial interface scaling usually analyzed through the interface width, also called roughness. The roughness is defined as the root mean square deviation of the interface around its mean value
\begin{equation}
W=\left[\frac{1}{N}\sum_{i=1}^N\left(r_i-\bar{r}\right)^2\right]^{1/2},
\label{eq:w}
\end{equation}
where a set of $N$ distances $r_i$ represents the interface and $\bar{r}$ is 
the corresponding mean value. The usual behavior of the roughness is a power law growth with the time $W \sim t^\beta$, where $\beta$ is the growth exponent. Intensive numerical simulations of the Eden model grown from flat substrates provide the growth exponent $\beta=1/3$ \cite{Vicsekbook,Kertesz,Devillard}, which corresponds to the Kardar-Parisi-Zhang (KPZ) universality class \cite{KPZ}. However, the roughness of on-lattice Eden aggregates grown from a seed is ruled by the lattice anisotropy \cite{Zabolitzky,Batchelor} and characterized by an asymptotic growth exponent $\beta=1$. Zabolitzky and Stauffer \cite{Zabolitzky} simulated clusters with $N\simeq 10^9$ particles of the square lattice Eden A model and observed that the interface width evolves in non-trivial way. In this version, the growth sites (empty sites with at least one occupied nearest-neighbor) are chosen at random and then occupied. For small clusters, a good agreement with KPZ growth exponent ($\beta\approx 1/3$) was observed, in contrast with the linear dependence on time ($\beta\rightarrow 1$) obtained for asymptotically large clusters. The value $\beta=1$ is associate to the cluster diamond-like shape caused by the square lattice anisotropy \cite{Batchelor1998}. 

Off-lattice Eden model was simulated by Wang \textit{et al.} \cite{Wang}, and the exponent $\beta=0.396$, a value significantly larger than the KPZ one, was found. Recently, Ferreira and Alves \cite{Edenoff} showed that the exponent reported by Wang \textit{et al.} is due to the fluctuations of the border center of mass  used as the origin for the roughness evaluation. Actually, the KPZ exponent is obtained when an origin fixed on the initial seed is used.

In the present work, we analyzed large Eden clusters ($N> 4\times 10^9$ particles) using a procedure developed in order to remove the anisotropy of DLA clusters \cite{Bogo,Alves}. Our central aim is to propose a strategy for the growth of isotropic on-lattice Eden clusters without change their universality class, as well as to corroborate recent finds about the scaling of off-lattice Eden clusters. The paper outline is the following. In Section \ref{Model}, the growth rules are presented and the results for DLA model summarized. The simulation results are presented and discussed in Section \ref{Results}, while some conclusions are drawn in Section \ref{Conclu}.

\section{\label{Model}Neighborhood dependent growth rule}

The growth of isotropic on-lattice DLA clusters was considered by Bogoyavlenskiy \cite{Bogo}. In this rule, an empty site is chosen to be grown following the standard DLA model, but the growth is implemented, or not, with probability $P_k$ given by
\begin{equation}
P_1:P_2:P_3\cdots P_k = 1^2:2^2:3^2\cdots k^2,
\end{equation}
where $k$ is the number of occupied nearest-neighbors of the growth site.  Although the preliminary simulations done by Bogoyavlenskiy suggested that the procedure was successful, Alves and Ferreira \cite{Alves} verified that it produces patterns with diagonal anisotropy at the large scale simulation or the high noise reduction limits. The noise reduction is a tool used to enhance the cluster anisotropy and consists of a set of counters at each growth site that are increased by a unit every time these sites are selected for the growth. An empty site is occupied only after its selection for $M$ times.
Also, Alves and Ferreira \cite{Alves} proposed a generalized algorithm with growth probabilities given by 
\begin{equation}
P_k=\left(\frac{k}{n}\right)^\nu,
\label{eq:pnu}
\end{equation}
where $n$ is the lattice coordination number and $\nu$ an adjustable parameter. It was found that a suitable choice of $\nu$ strongly reduces the cluster anisotropy. Indeed, the DLA patterns exhibit axial and diagonal anisotropies when $\nu<\nu_c$ and $\nu>\nu_c$, respectively, and an eight-fold structure was observed for $\nu=\nu_c$. In the generalized Eden model, an empty site is chosen to be grown using one of the Eden model versions and the growth is implemented, or not, following Eq. (\ref{eq:pnu}). 

\section{\label{Results}Simulations}
\begin{figure*}[hbt]
\begin{center}
\includegraphics[width=13.5cm,height=!,clip=true]{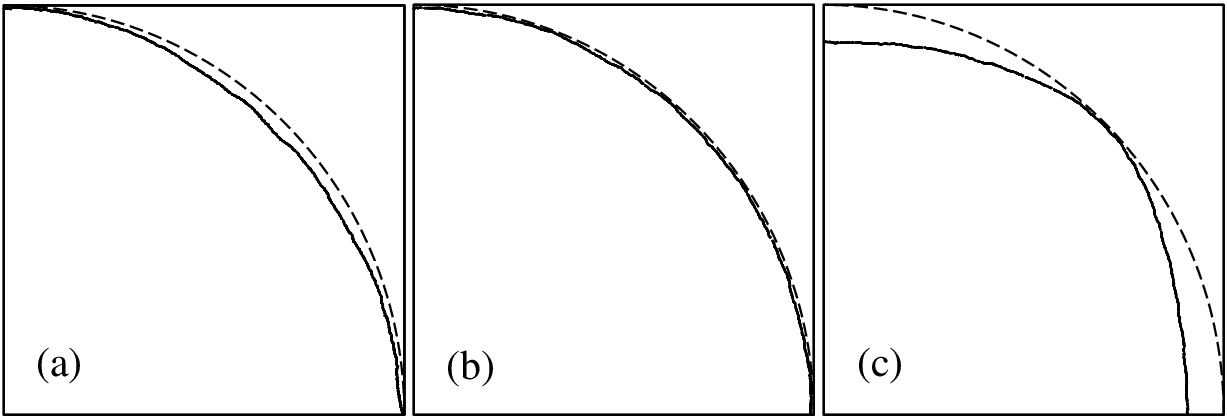}\\
\end{center}
\caption{\label{fig:m1} Contours of large Eden B clusters (only one quadrant is shown) without noise reduction for (a) $\nu=0$ (original model), (b) $\nu=1.0$, and (c) $\nu=2.0$.  In all patterns, the system size is $L=10^4$ and the total number of particles is $N\sim10^8$. The dashed lines represent circles of diameter $L$ used to illustrate the deviations from the circular shape.} 
\end{figure*}

In this section, the square lattice case with lattice coordination $n=4$ is considered. Figure \ref{fig:m1} shows the borders of large Eden B clusters (linear size $L=10^4$) generated without noise reduction ($M=1$) in order to illustrate the deviations from the circular shape. For sake of comparison, circles of diameter $L$ centered at the seed were drawn (dashed line). It is evident from this figure that the border is axially or diagonally stretched for small and large $\nu$ values, respectively, whereas an approximately symmetric shape is observed when $\nu_{c}\approx 1$. This $\nu_c$ value differs from the value $\nu_c=0.507\pm 0.005$ found for the noiseless limit as observed for the DLA model \cite{Alves} where the crossover occurs at $\nu_c=1.395\pm0.005$ and $\nu_c=3.4\pm0.2$ for the noiseless and $M=1$ cases, respectively.

\begin{figure*}[hbt]
\begin{center}
\subfigure[]{\includegraphics[clip=true,width=6.8cm,height=!]{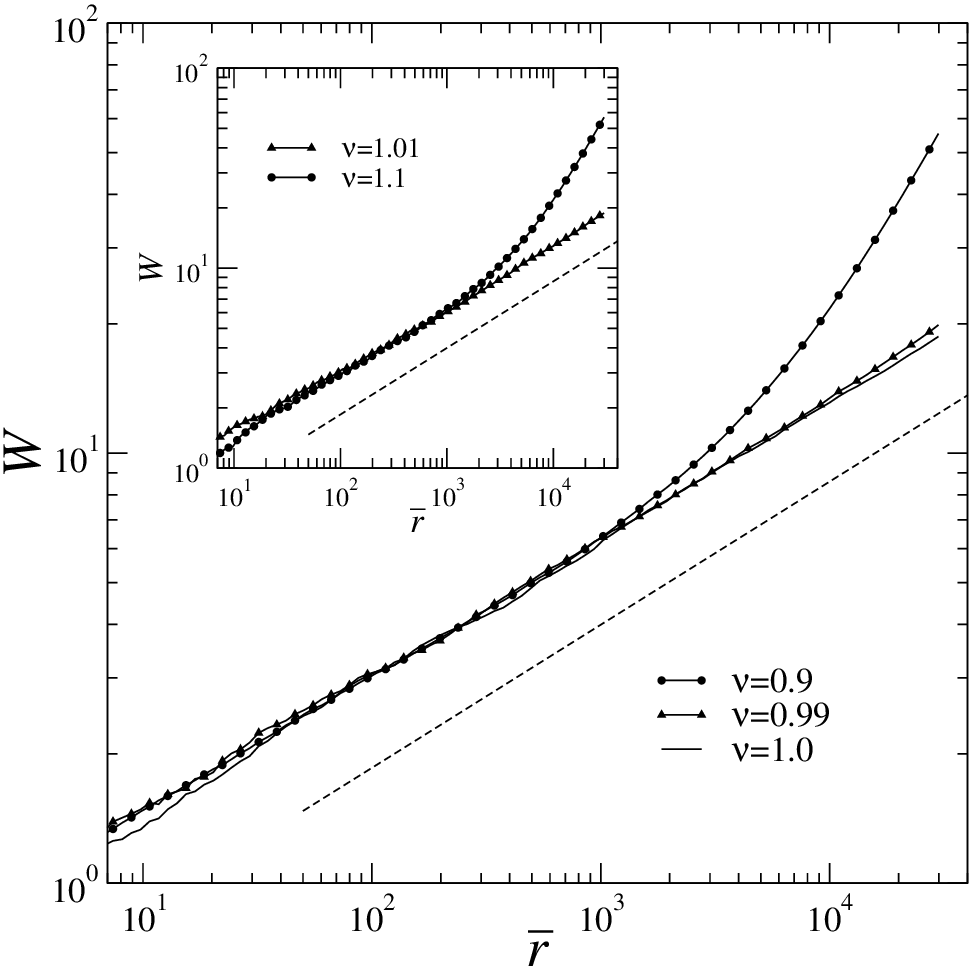}}
\subfigure[]{\includegraphics[clip=true,width=6.8cm,height=!]{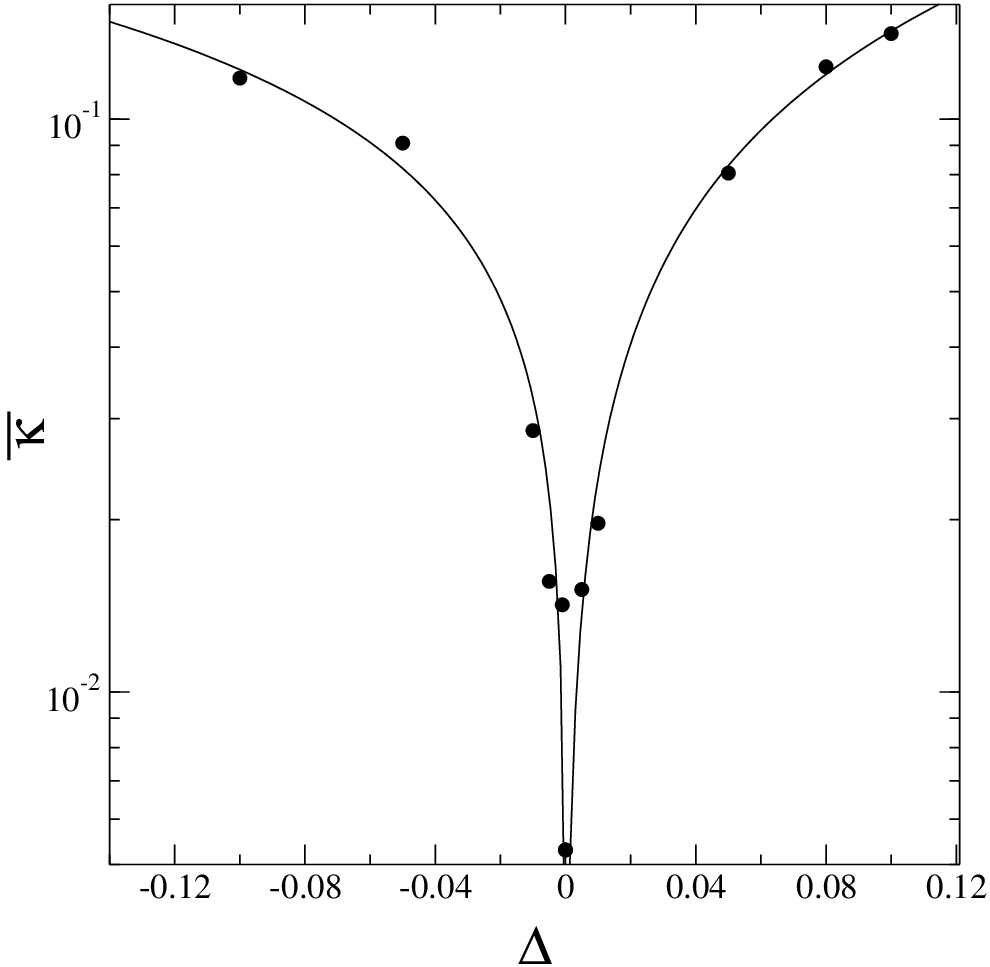}}
\end{center}

\caption{\label{fig:wedenc} (a) Interface width as a function of the cluster mean radius for distinct $\nu$ values and (b) mean curvature as a function of $\Delta=\nu-\nu_c$. In (a), the main figure shows the curves for $\nu\le 1$ while the inset shows those for $\nu>1$. The dashed lines represent power law $w \sim t^{\frac{1}{3}}$. In (b), the solid lines are the power laws $\bar{\kappa}\sim (-\Delta)^{0.57}$ and $\bar{\kappa}\sim \Delta^{0.78}$ for $\Delta<0$ and $\Delta>0$, respectively.  In all data, the averages were done over $120$ independent samples.} 
\end{figure*}

Two central points should be elucidated. The first one concerns the quantitative estimate for $\nu_c$, and the second one is the inquiry if the model belongs to the KPZ universality class for $\nu=\nu_c$. In order to analyze these points, we determined the interface width evolution considering the center of the lattice as the origin. Since the model dynamics is restricted to the border, we can do an analogy with the growth from flat substrates \cite{Vicsekbook} and take the time proportional to the number of peripheral particles or, equivalently, to the mean radius $\bar{r}$. We used square lattices containing $64000\times 64000$ sites and grew clusters with more than $4\times 10^9$ particles. We concentrate our analyzes on the Eden B model because its convergence to the asymptotic scaling is faster than that of the Eden A model \cite{Meakinbook}. For the last one, $N\approx 10^9$ particles were necessary to enhance the diamond-like scaling ($w\sim t$) \cite{Zabolitzky}. In Figure \ref{fig:wedenc}(a), the interface width is plotted as a function of the mean radius for $\nu$ varying around $\nu=1$. As one can see, the double-logarithm curves exhibit upward curvatures for both $\nu<1.0$ and $\nu>1.0$ (inset) while a power law is observed for $\nu\approx 1$. Also, the slopes are very close to the KPZ growth exponent $\beta=1/3$ when $\nu\approx 1$. 

In order to find the null curvature point, which represents the absence of anisotropy, the double-logarithm data are fitted by cubic polynomials $P_3(r)$ and the mean curvature is defined by
\begin{equation}
\bar{\kappa}=\frac{1}{\log_{10} r_f-\log_{10} r_i}\int_{\log_{10}  r_i}^{\log_{10}  r_f} \kappa(r)dr,
\end{equation}
where $r_i$ and $r_f$ are the lower and upper cutoffs of the data used in the cubic fits, respectively, and $\kappa(r)$ is the local curvature defined by
\begin{equation}
\kappa(r)=\frac{P_3''}{[1+(P_3')^2]^{3/2}}.
\end{equation} 
The transient behavior for very small clusters was avoided using fit ranges from $r_i \approx 46$ to $r_f \approx 3\times10^4$. The mean curvature as a function of $\nu$ is plotted in figure \ref{fig:wedenc}(b). This figure suggests a abrupt variation around the minimum of $\bar{\kappa}$. The minimum of $\bar{\kappa}$ occurs between $\nu=1.001$ and $\nu=0.999$ providing the estimates $\nu_c=1.000\pm0.001$ and \textbf{$\beta=0.325\pm 0.017$}. Here, the fits were done for $\bar{r}>100$ and the uncertainty represents the local slope fluctuations along the distinct parts of the curves. This growth exponent is in very good agreement with the KPZ universality class expected for the original Eden model. Finally, these data strongly indicate that the curvature vanishes as a power law as $\nu\rightarrow \nu_c$.

\begin{figure*}[ht]
\begin{center}
\includegraphics[width=8cm,height=!,clip=true]{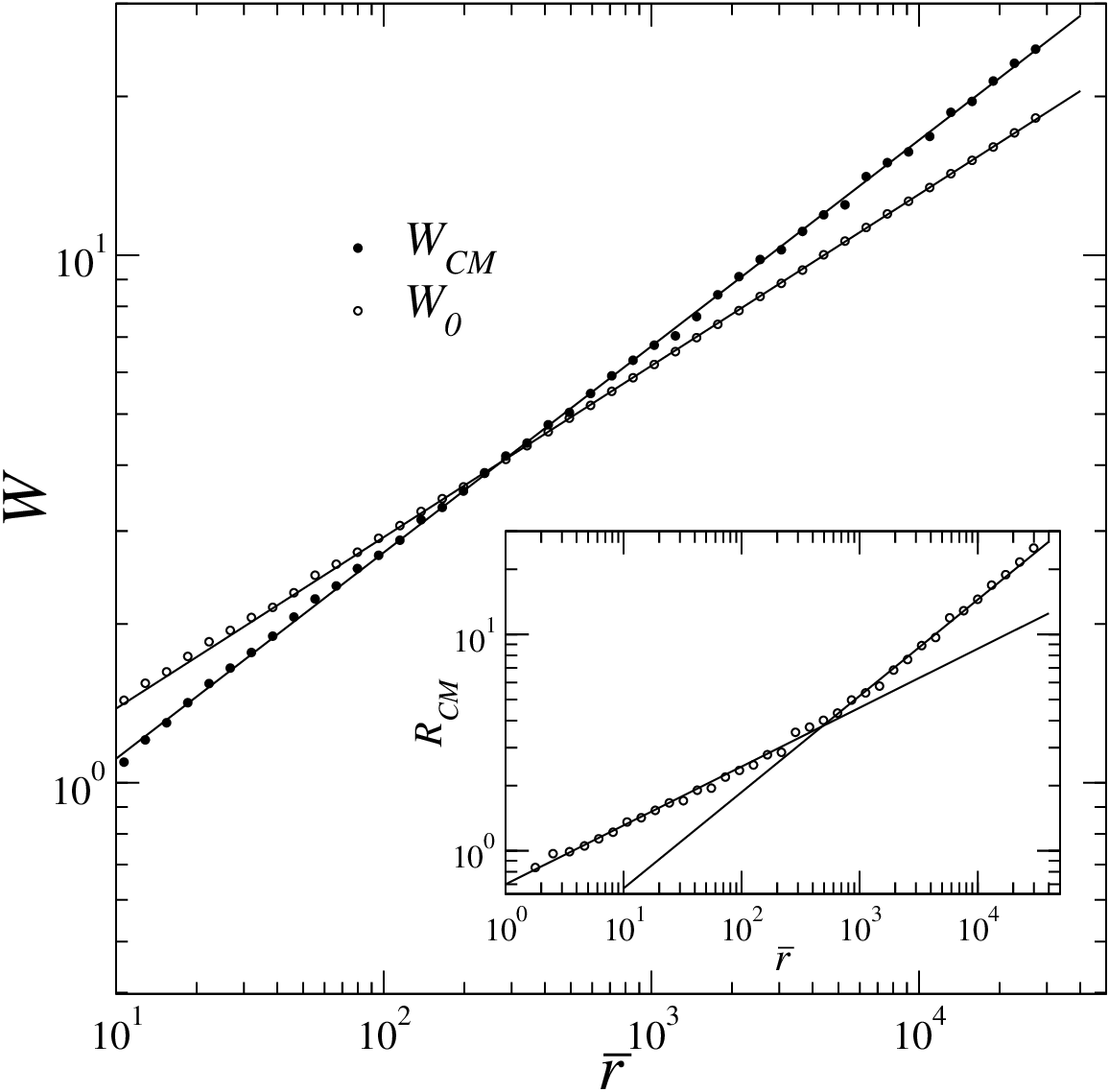}
\end{center}
\caption{\label{fig:wcm} Interface width (two methods) and CM mean distance from the initial seed (inset) as functions of the mean radius for $\nu=1$. The straight lines are power law fits. The averages were done over $120$ independent samples.}
\end{figure*}

A central feature related to the interface scaling of round clusters is the choice of the origin, as recently demonstrated for the off-lattice Eden model \cite{Edenoff}. In particular, if the width is evaluated in relation to the border center of mass (CM), the growth exponent is $\beta_{CM}\simeq 2/5$ \cite{Edenoff}, which differs from the KPZ exponent observed when the initial seed is used (a fixed origin). This difference was associated to the CM fluctuations growing faster than the interface ones. We applied this analysis to the present model with $\nu=\nu_c$ in order to assess the agreement with the scaling exponents obtained for the off-lattice model and to extend the off-lattice simulations limited to $3\times10^7$ particles to more than $4\times 10^9$ particles. The results are show in Figure \ref{fig:wcm}, where the interface width evaluated in relation to the CM ($W_{CM}$) is compared with that in relation to the seed ($W_0$). It is evident that the corresponding growth exponents are distinct. Indeed, the exponents are $\beta_0=0.324\pm0.017$ and $\beta_{CM}=0.392\pm0.016$. These results are in excellent agreement with the off-lattice Eden model \cite{Alves}, reinforcing the equivalence between the universality classes of the models. The CM mean distance from the seed exhibits two power law regimes, namely $R_{CM}\sim \bar{r}^{~0.26}$ and $R_{CM}\sim \bar{r}^{~0.46}$ for small and large system sizes, respectively. Again, the asymptotic scaling agrees with the off-lattice simulations.

The most used version of the Eden model is that known as Eden A \cite{Meakinbook}. In this version a growth site (an empty site neighboring an occupied one) is chosen at random and then occupied. In the noiseless limit, the transition from axial to diagonal anisotropy is the same observed for the Eden B model, including the interface width minimum occurring at $\nu_c \thickapprox 0.51$. Without noise reduction, the criteria of null curvature gives $\nu_c=1.72\pm0.01$ and growth exponents in very good agreement with the off-lattice Eden model. The lower precision of the parameter $\nu_c$, in comparison with the Eden B version, is due to the slower convergence to the asymptotic exponents and the larger $\nu_c$ value. The last implies in a higher frequency of rejected tentatives along the simulations.

\begin{figure*}[ht]
\begin{center}
\includegraphics[width=8cm,height=!,clip=true]{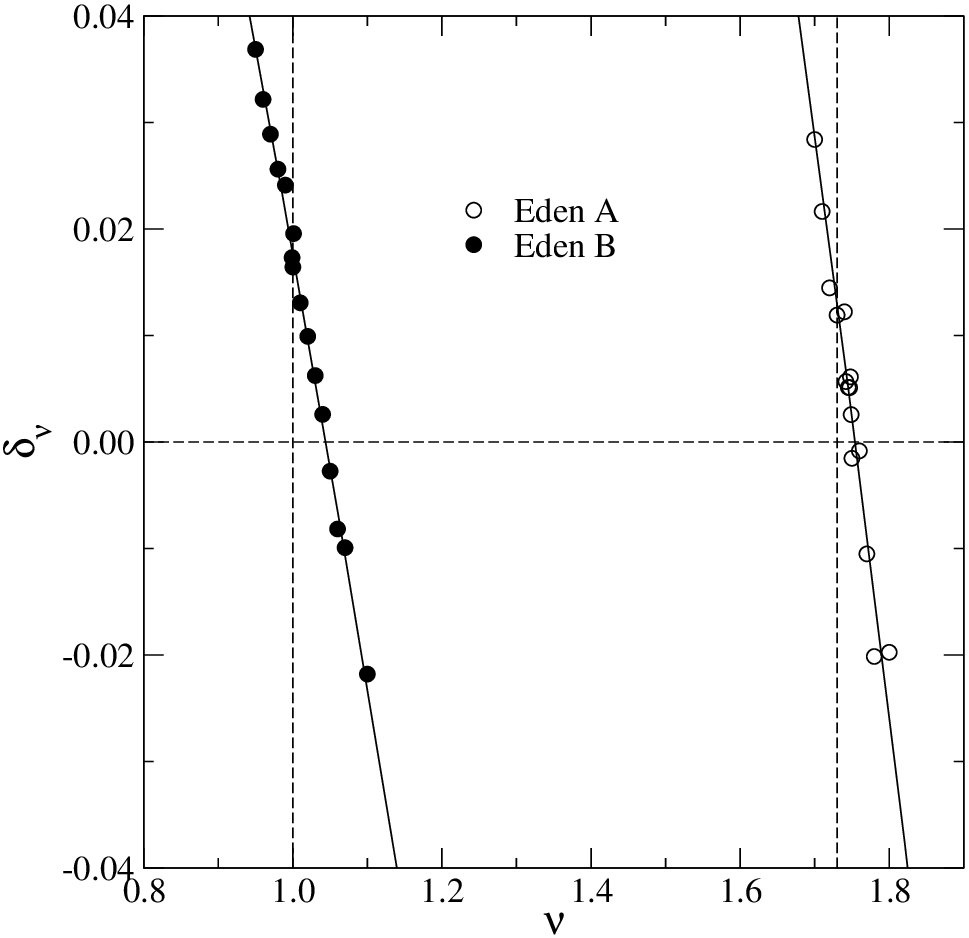} \\ 
\end{center}
\caption{\label{fig:delta} Differences between diagonal and axial growth. The dashed vertical lines indicates the $\nu_c$ values for Eden A (right) and B (left) models. The averages were done over $1000$ samples of size $L=400$. It is important to stress that the histograms remains essentially unchanged as the system size is varied.}
\end{figure*}

The difference between the $\nu_c$ values for Eden A and B models can be understood  analyzing the growth site neighborhood at $\nu\approx\nu_c$. Lets $Q_k$ be the fraction of growth sites with $k$ occupied neighbors. Considering the square lattice anisotropy, one can easily conclude that the growth sites with $k=1$ and $k=3$ produce axial growth, those with $k=2$ produce diagonal growth, while those with $k=4$ produce an inward growth which does not contribute for the overall anisotropy.  One can suppose that if the axial growth probability exceeds the axial one, the anisotropy rotation emerges. So, we analyzed the difference between them using the relation
\begin{equation}
\delta_\nu=4^\nu(P_{axial}-P_{diagonal})=Q_1(\nu) + 3^\nu Q_3(\nu) -2^\nu Q_2(\nu).
\end{equation}
Figure \ref{fig:delta} shows the curves $\delta_\nu$ against $\nu$. As one can see, the $\nu=\nu^\star$ values for which the diagonal and axial growth probabilities are equal ($\delta_\nu=0$) are slightly larger than $\nu_c$. Indeed, we found $\nu^\star-\nu_c\simeq0.02$ and $\nu^\star-\nu_c\simeq0.04$ for the Eden A and B, respectively.
Notice that these differences are small, but neatly distinguishable since the respective deviations from null curvature are evident as shown in Figure \ref{fig:wedenc}(b). Thus, we conclude that the isotropic growth of Eden clusters cannot be explained only by the counterbalance between diagonal and axial growths.

\section{\label{Conclu} Conclusions}

In this work, we developed a strategy to grow on-lattice Eden clusters without the undesirable anisotropy effects. Asymptotically large clusters generated with the original Eden model are ruled by a diamond-like and, consequently, a growth exponent $\beta=1$, in disagreement with the KPZ universality class ($\beta=1/3$) observed for the growth from a flat substrate. We proposed an extended Eden model, in which the growth probability is proportional to a power of the number of occupied nearest neighbors, $P_k\propto k^\nu$, in analogy with a procedure developed to remove the anisotropy of DLA clusters \cite{Bogo,Alves}. We found that the choices $\nu=1.72 \pm 0.01$ and $\nu=1.000\pm 0.001$ lead to isotropic patterns for the Eden A and B models, respectively, preserving the KPZ universality class. Also, the extended model agrees with off-lattice simulations, in which the center of mass fluctuations are considered \cite{Edenoff}.

The present strategy constitutes a useful tool for modeling experiments with radial symmetry since lattice models can be easily generalized. A very important example is the recent report concerning the dynamics of tumors \cite{Bru1998,Bru2003}. These experiments indicate a universal dynamics characterized by the Mullins-Herring universality class \cite{MH} for distinct types of tumors. However, this idea is subject of recent controversies \cite{Buceta}.

\bigskip
\noindent \textbf{Acknowledgments}\\
~\\
The work was supported by the Brazilian Agencies CNPq, FAPEMIG, and CAPES.

\end{document}